\documentclass[english,11pt,aps,amsfonts,amssymb,superscriptaddress]{revtex4}
\usepackage{amsmath}
\usepackage{babel}
\usepackage{bm}
\usepackage{enumerate}

\begin{document}

\newcommand\new[1]{\ensuremath{\blacktriangleright}#1\ensuremath{\blacktriangleleft}}
\newcommand\note[1]{$\blacktriangleright$[\emph{#1}]$\blacktriangleleft$}
\newcommand\nw[1]{\ensuremath{\spadesuit}#1\ensuremath{\spadesuit}}

%%%%%%%%%%%%%%%%%%%%%%%
\title{Massive motion in Brans--Dicke geometry and beyond}

\author{Raffaele Punzi}
\email{raffaele.punzi@desy.de}
\affiliation{Zentrum f\"ur Mathematische Physik und II. Institut f\"ur Theoretische Physik, Universit\"at Hamburg, Luruper Chaussee 149, 22761 Hamburg, Germany}

\author{Frederic P. Schuller}
\email{f.p.schuller@nucleares.unam.mx}
\affiliation{Max Planck Institut f\"ur Gravitationsphysik, Albert Einstein Institut, Am M\"uhlenberg 1, 14467 Potsdam, Germany}

\author{Mattias N.\,R. Wohlfarth}
\email{mattias.wohlfarth@desy.de}
\affiliation{Zentrum f\"ur Mathematische Physik und II. Institut f\"ur Theoretische Physik, Universit\"at Hamburg, Luruper Chaussee 149, 22761 Hamburg, Germany}

%%%%%%%%%%%%%%%%%%%%
\begin{abstract}
Gravity theories that can be viewed as dynamics for area metric manifolds, for which Brans--Dicke theory presents a recently studied example, require for their physical interpretation the identification of the distinguished curves that serve as the trajectories of light and massive matter. Complementing previous results on the propagation of light, we study effective massive point particle motion. We show that the relevant geometrical structure is a special Finsler norm determined by the area metric, and that massive point particles follow Finsler geodesics.
\end{abstract}
\maketitle

\numberwithin{paragraph}{section}

%%%%%%%%%%%%%%%%%%%%%%%%%%%%%%%%%%%%%%%%%%%%%%%%%%%%%%%%%%%%%%%%%%%%%%%%%%%%%%
\section{Introduction}
The non-metric geometry underlying Brans--Dicke theory~\cite{Brans:1961sx} was revealed in a recent letter~\cite{Punzi:2008dv}. The metric and scalar field were unified into the single geometric structure of an area metric~\cite{Schuller:2005ru}, and the Brans--Dicke equations for vanishing parameter $\omega$ were those of the area metric refinement of Einstein--Hilbert gravity~\cite{Punzi:2006nx}. As far as the vacuum equations are concerned, the area metric interpretation is fully equivalent to the scalar tensor interpretation. 

But the inclusion of matter leads to a surprise. As we will show in this article, the motion of massive particles in area metric spacetime does not follow metric geodesics, but special Finsler geodesics. The present discussion of massive matter complements previous results on the motion of light rays~\cite{Punzi:2007di}, which remarkably is governed by the same Finsler geometry. We emphasize that these results are valid for any area metric spacetime. 

For the particular area metric identified in the context of $\omega=0$ Brans--Dicke theory, the non-geodesic motion renders the theory consistent with solar system physics to post-Newtonian order~\cite{Punzi:2008dv}. This is not the case if matter is coupled only to the metric~\cite{Will:2005va}, and the special contribution of the scalar field dictated by area metric geometry is ignored. The idea of employing the scalar field to modify the metric matter coupling in scalar tensor theories is not new~\cite{Magnano:1993bd,Deruelle:2008fs}; but the area metric point of view uniquely links the $\omega=0$ theory to a distinguished and consistent matter coupling.

Quite generally area metric geometry presents a way to extend metric gravity theories beyond simple scalar tensor theories. Our results on the motion of massive particles presented here and the motion of light rays are essential to discuss the phenomenology of area metric spacetimes, and thus the physical consistency of any particular gravitational dynamics.

In the technical part of this article, we will determine a class of distinguished curves which are associated with massive motion on area metric manifolds. This is possible despite the fact that area metrics a priori do not define a natural length measure. Neither do they admit the standard notion of a perfect fluid. But there exist refined fluids whose fundamental mechanical constituents can be thought of as classical strings; this is discussed in section~\ref{sec_fluid}.  In section~\ref{sec_effective}, we will identify interaction terms that cause the strings to clump together so that they effectively behave as a particle fluid. Our central result is that the motion of the effective particles follows the geodesics of a special Finsler norm determined by the area metric.  The consistency of this result with any diffeomorphism invariant gravity action for area metric backgrounds is shown in section~\ref{sec_consistency}. We conclude with a discussion in section~\ref{sec_discussion}.

%%%%%%%%%%%%%%%%%%%%%%%%%%%%%%%%%%%%%%%%%%%%%%%%%%%%%%%%%%%%%%%%%%%%%%%%%%%%%
\section{String fluids}\label{sec_fluid}
Perfect fluids on area metric backgrounds feature a refined structure which can be viewed as arising from strings rather than point particles being their constitutive matter. In this section we briefly review and elaborate on some known results on string fluids, based on a precise definition of area metric geometry. 

Area metric manifolds $(M,G)$ feature a smooth covariant fourth-rank tensor field $G_{abcd}$ with the symmetries $G_{abcd}=G_{cdab}$ and $G_{abcd}=-G_{bacd}$. Further the area metric is required to be invertible in the sense that there exists a smooth contravariant fourth-rank tensor field $G^{abcd}$ so that $G^{abpq}G_{pqcd}=4\delta^{[a}_c\delta^{b]}_d$. All matter on area metric backgrounds is described by a fourth rank source tensor $T^{abcd}$ which arises from the variation of the matter action with respect to the area metric, and is defined in \cite{Punzi:2006nx}. Important for the present paper is the observation that independent of any assumed gravitational dynamics, diffeomorphism invariance of the matter action implies a conservation equation for the source tensor, which can be written in the form
\begin{equation}\label{conservation}
-|\textrm{Det }G|^{1/6}T^{ijkl}\partial_pG_{ijkl}+4\partial_a\left(|\textrm{Det }G|^{1/6}T^{ijka}G_{ijkp}\right)=0\,.
\end{equation}

We now consider a particular form of matter on area metric backgrounds, namely string fluids~\cite{Punzi:2006nx,Punzi:2006hy,Schuller:2007ix}. As on metric backgrounds~\cite{Letelier:1979ej,Stachel:1980zr,Stachel:1980zs,Letelier:1984dm}, these can be thought of as collections of strings. Geometrically, their description features a field of local tangent areas $\Omega=\dot x\wedge x'$, i.e., ${\Omega^{ij}=\dot x^ix'^j-\dot x^jx'^i}$, to the two-dimensional string worldsheets ${x:\Sigma\rightarrow M}$. This is analogous to perfect fluids in general relativity, which can be understood as a collection of point particles, and whose description involves the velocity field tangent to the particle worldlines. Even though string fluids are not derived from an action, their source tensor must satisfy the conservation equation above in order to ensure a consistent coupling to any theory of area metric gravity which is derived from an action by variation with respect to the area metric $G$.

The simplest string fluid is non-interacting string dust with source tensor
\begin{equation}\label{dustsource}
T^{ijkl}=\tilde\rho \Omega^{ij}\Omega^{kl}\,.
\end{equation}
That this indeed describes non-interacting strings will now be shown by proving that the source conservation equation is equivalent to the equation of motion of the free classical string, i.e., the minimal surface equation, and the string continuity equation. 

To see this, consider the string worldsheet tangent areas to be normalized as $G(\Omega,\Omega)=-1$ for $\Omega=u\wedge v$, where $u=\dot x$, $v=x'$. Substituting the source tensor~(\ref{dustsource}) into the conservation equation~(\ref{conservation}), one obtains
\begin{eqnarray}\label{dustcon}
0 & = & |\textrm{Det }G|^{1/6}\tilde\rho\,\Big\{v^q\partial_q(G_{apcd}u^au^cv^d)+u^q\partial_q(G_{pbcd}v^bu^cv^d)-\frac{1}{2}\partial_pG_{abcd}u^av^bu^cv^d\Big\}\nonumber\\
&& \,+\,\,G_{ijkp}u^iv^j\Big[\partial_l\left(|\textrm{Det }G|^{1/6}\tilde\rho\Omega^{kl}\right)\!\Big].
\end{eqnarray}
The minimal surface equation for strings on area metric backgrounds is derived as the stationarity condition of the integrated worldsheet area~\cite{Schuller:2005ru}, and requires the vanishing of the curly brackets in the expression above. The continuity equation~\cite{Punzi:2006nx} on the other hand requires the vanishing of the square brackets. Hence both these conditions together imply source conservation; this direction of the argument was already given in~\cite{Punzi:2006nx}.

Now to show also the converse, observe that the first line in the equation above vanishes if contracted with $\Omega^{pq}$; hence the second line does. With the notation $\Omega_{pq}=G_{pqrs}\Omega^{rs}$ one thus concludes
\begin{equation}\label{condi}
\Sigma_{mk}\partial_l\left(|\textrm{Det }G|^{1/6}\tilde\rho\Omega^{kl}\right)=0
\end{equation}
for $\Sigma_{mk}=\Omega_{mq}\Omega^{qp}\Omega_{pk}$. The next step is to show that $\Sigma_{mk}$  in this equation can be replaced simply by $\Omega_{mk}$. This follows from the easily checked identity $\Omega^{ab}\Omega_{bc}\Omega^{cd}=\Omega^{ad}$, whence $\Sigma_{mk}\Omega^{kl}=\Omega_{mk}\Omega^{kl}$. By linear independence of the vectors $u$ and $v$ (otherwise the tangent area $\Omega=0$ would be degenerate), it follows that
$\Sigma_{mk}u^k=\Omega_{mk}u^k$ and ${\Sigma_{mk}v^k=\Omega_{mk}v^k}$, i.e., that $\Sigma_{mk}$ can be replaced by $\Omega_{mk}$ in contractions with $u$ and $v$. This is precisely what we need for the replacement of $\Sigma$ by $\Omega$ in (\ref{condi}), and so we see from (\ref{dustcon}) that source conservation implies the minimal surface condition for string dust. Finally, we may rewrite~(\ref{condi}) as
\begin{equation}
G(u,v,u,\cdot)\,\partial_l\left(|\textrm{Det }G|^{1/6}\tilde\rho v^l\right)-G(u,v,v,\cdot)\,\partial_l\left(|\textrm{Det }G|^{1/6}\tilde\rho u^l\right)=0\,.
\end{equation}
Evaluating this one-form on $u$ and $v$, respectively, shows that both divergence terms must vanish separately. Hence also the continuity equation,
in the form of vanishing square brackets in~(\ref{dustcon}), holds.

With this new converse result, it is now rigorously proven that non-interacting string dust on area metric backgrounds is described by the source tensor given in~(\ref{dustsource}). Any modification of this source tensor by other terms depending on the background geometry~$G$ or the worldsheet tangent areas~$\Omega$, hence describes an interacting string fluid. 

Finally consider the familiar case of a purely metric spacetime $(M,g)$ where areas are simply measured by the induced area metric $G_{abcd}=2g_{a[c}g_{d]b}$. Then the source conservation equation~(\ref{conservation}) for string dust with source tensor~(\ref{dustsource}) reduces to $\nabla_a T_\textrm{eff}{}^a{}_b=0$ for $T_\textrm{eff}{}^a{}_b=\tilde\rho\Omega^{ab}\Omega_{pb}$. Writing $\Omega=u\wedge v$ as we did before, and choosing the basis $g(u,v)=0$, $g(u,u)=-1$ and $g(v,v)=1$, one obtains
\begin{equation}\label{metriclimit}
T_\textrm{eff}{}^a{}_b=\tilde\rho (u^a u_b-v^a v_b)
\end{equation}
for string dust. This special case of our construction is known from the literature, and has been used to describe string energy momentum coupled to standard metric theories of gravity \cite{Letelier:1979ej,Stachel:1980zr}.

%%%%%%%%%%%%%%%%%%%%%%%%%%%%%%%%%%%%%%%%%%%%%%%%%%%%%%%%%%%%%%%%%%%%%%%%%%%%
\section{Effective Finsler geodesics}\label{sec_effective}
We are now prepared to derive the key result of this paper. In this section we will demonstrate the existence of a class of interacting string fluids that behave precisely the same way as non-interacting particle dust. We prove that these string fluids effectively propagate along non-null Finsler geodesics with respect to a special Finsler norm determined by the area metric. 

Consider again the case of non-interacting strings on a metric background with effective energy momentum~(\ref{metriclimit}). It is clear that such energy momentum cannot be interpreted as that of a point particle fluid: the string worldsheet singles out a preferred spatial direction $v$ which destroys isotropy around the particle trajectory $u$. The term $\tilde\rho v^av_b$ represents anisotropic pressure. From this observation, which is not new for metric backgrounds, we learn the following lesson for area metric geometry. Two issues must be taken into account to derive string fluids which effectively behave like non-interacting point particle fluids: we must
\begin{itemize}
\item isotropically superpose string fluids by implementing an average over the spatial directions of the respective worldsheets; and
\item adjust the string interaction terms so to achieve effective point particle motion. 
\end{itemize}
This is the procedure we will now implement. In~\ref{sec_isodust}, we will define the isotropic average of string dust matter; in~\ref{sec_mattra}, we then determine the necessary string interaction terms responsible for non-interacting particle motion. The fact that interaction terms have to be added to an isotropic average of non-interacting strings agrees with the physical intuition that strings, in order to effectively behave like point particle dust, must clump together by some form of interaction.

\subsection{Isotropization of string dust}\label{sec_isodust}
As discussed above, the first step in finding string fluids that effectively move as non-interacting particle fluids is the definition of an isotropic average over the spatial worldsheet directions. We define this average with respect to a vector field $u$ which later emerges as the velocity field of the resulting particle fluid. The construction will be independent of coordinates. 

Note first that the local tangent spaces can be decomposed as $TM=\left<u\right>\oplus V$ (the following construction will not be affected by the non-uniqueness of the complement $V$), which in turn induces a decomposition of the antisymmetric tensor bundle as $\Lambda^2TM=\Lambda^2_uTM\oplus \Lambda^2_VTM$ for
\begin{equation}
\Lambda^2_uTM=\left\{\Omega\in \Lambda^2TM\,|\,\Omega\wedge u=0\right\}.
\end{equation}
Thus any element of the space $\Lambda^2_uTM$ can be written in the form $\Omega=u\wedge v$ for $v\in V$, but $\Lambda^2_uTM$ is independent of the choice of complement $V$, since $u\wedge v=u\wedge (v+\lambda u)$ for any scalar $\lambda$. Moreover, $\Lambda^2_uTM$ is a three-dimensional linear subspace of $\Lambda^2TM$ to which the area metric hence can be sensibly restricted. The restriction then defines a unique metric $\tilde g:\Lambda^2_uTM\times \Lambda^2_uTM\rightarrow\mathbb{R}$ by
\begin{equation}
\tilde g=-G|_{\Lambda^2_uTM}
\end{equation}
on all areas in the set $u\wedge V$ independent of the choice of $V$. We assume that $\tilde g$ is positive definite; note that this is not a restriction on the background geometry, but distinguishes particular vector fields $u$ that can play the role of velocity field in our final particle fluid. 

The sought-for isotropic average over the spatial worldsheet directions will now essentially be the integration over the linear subspace $\Lambda^2_uTM$ with metric measure $\tilde g$. But since the volume of this space is non-compact, we must restrict the integration to the two-dimensional unit sphere~$S^2_u$ consisting of areas with $\tilde g(\Omega,\Omega)=1$. Let $\phi^*$ denote the pullback from $\Lambda^2_uTM$ to $S^2_u$ and use coordinates $\theta^1,\theta^2$. Then $\textrm{vol }S^2_u=\int d^2\theta\,\sqrt{\textrm{det }\phi^*\tilde g}=4\pi$, which is most easily seen by choosing the Cartesian frame $\{e_{\hat 0},e_{\hat\alpha}\}$ with $e_{\hat 0}=u$ and $\left<e_{\hat\alpha}\right>=V$ such that $\tilde g_{\hat\alpha\hat\beta}=\tilde g(e_{\hat 0}\wedge e_{\hat\alpha},e_{\hat 0}\wedge e_{\hat\beta})=\delta_{\hat\alpha\hat\beta}$. We now calculate the average of the string dust source tensor~(\ref{dustsource}) over $S^2_u$,
\begin{equation}
   \big<T^{abcd}\big>    = \frac{\tilde\rho}{\textrm{vol }S^2_u}\int_{S^2_u}d^2\theta\,\sqrt{\textrm{det }\phi^*\tilde g}\,\Omega(\theta)^{ab}\Omega(\theta)^{cd}\,.
\end{equation}
The integral is performed using the same Cartesian frame as above; for elements of $S^2_u$ we then have $\Omega=\Omega^{\hat\alpha}e_{\hat 0}\wedge e_{\hat\alpha}/\sqrt{\tilde g(\Omega,\Omega)}$ and hence the coordinates $\Omega^{\hat\alpha}/||\Omega||$. As a result we find
\begin{equation}
\big<T^{\hat 0\hat\alpha\hat 0\hat\beta}\big> = \frac{1}{3}\tilde\rho\delta^{\hat\alpha\hat\beta}\,.
\end{equation}
Note that in this frame $\delta^{\hat\alpha\hat\beta}=\tilde g^{\hat\alpha\hat\beta}$ which is defined as the inverse of $\tilde g_{\hat\alpha\hat\beta}$ regarded as a $3\times 3$ matrix:
\begin{equation}
\tilde g^{\hat\alpha\hat\beta} = \frac{1}{2\textrm{det }\tilde g}\epsilon^{\hat 0\hat\alpha\hat\mu\hat\nu}\epsilon^{\hat 0\hat\beta\hat\rho\hat\sigma}\tilde g_{\hat\mu\hat\rho}\tilde g_{\hat\nu\hat\sigma}\,.
\end{equation}
Since the areas over which we averaged are elements of $\Lambda^2_uTM$, the result of the average, which is a linear operation, must be a tensor $\Lambda^{2\,*}_uTM\times\Lambda^{2\,*}_uTM\rightarrow \mathbb{R}$. Hence there is an extension $\tilde g^{-1}:\Lambda^{2\,*}_uTM\times\Lambda^{2\,*}_uTM\rightarrow \mathbb{R}$ with components $\tilde g^{-1\,abcd}$ that in the frame chosen above reduces to $\tilde g^{-1\,\hat 0\hat\alpha\hat 0\hat\beta}=\tilde g^{\hat\alpha\hat\beta}$. In other words, this allows us to write the result of the average in the fully covariant form
\begin{equation}\label{Tiso}
\big<T^{abcd}\big>
= \frac{1}{3}\tilde\rho \tilde g^{-1\,abcd}
= \frac{1}{3}\tilde\rho \frac{1}{2}u^{[a}u^{[c}h^{d]b]}
\end{equation}
with
\begin{eqnarray}
h^{ab} & = & \mathcal{G}(u,u,u,u)^{-1}\omega_{G^C}^{armn}\omega_{G^C}^{bspq}G^C_{rmtp}G^C_{vnsq}u^tu^v\,,\\
\mathcal{G}_{abcd} & = & -\frac{1}{24}\omega_{G^C}^{ijkl}\omega_{G^C}^{mnpq}G^C_{ijm(a}G^C_{b|kn|c}G^C_{d)lpq}\,.\label{dualFresnel}
\end{eqnarray}
The antisymmetrizations in $\tilde g^{-1}$ act only on the index pairs, $G^C$ denotes the cyclic part ${G_{abcd}-G_{[abcd]}}$ of the area metric, and the volume form is $\omega_{G^C}^{abcd}=|\textrm{Det }G^C|^{-1/6}\epsilon^{abcd}$, where $\textrm{Det}$ is the determinant taken over $G^C:\Lambda^2TM\times\Lambda^2TM\rightarrow\mathbb{R}$ regarded as a $6\times 6$ matrix by considering its antisymmetric index pairs.

We remark that the equations that could now be obtained from the conservation of the isotropic averaged source tensor (\ref{Tiso}) only involve the vector field $u$ and the background geometry determined by the area metric $G$, so they are already equations for a particle fluid, albeit an interacting one.

\subsection{Matter trajectories}\label{sec_mattra}        
We now come to the second part of the programme for this section, and determine the necessary string interaction terms that have to be added to the isotropic averaged source tensor (\ref{Tiso}) so that the resulting string fluid moves as a non-interacting particle fluid.

It will turn out to be sufficient to consider interaction terms $\Sigma(G)^{abcd}$ that only depend locally on the background geometry. Our ansatz for the particle string fluid source tensor therefore is
\begin{equation}\label{sourceps}
T^{abcd}= \frac{1}{3}\tilde\rho \tilde g^{-1\,abcd} + \frac{4}{3}\tilde\rho \Sigma^{abcd}\,.
\end{equation}
We will now determine the term $\Sigma$ so that the source conservation equation (\ref{conservation}) implies the standard continuity equation for point particles,
\begin{equation}\label{continuity}
   \partial_l\left(|\textrm{Det }G|^{1/6}\tilde\rho \tilde A u^l\right)=0  
\end{equation}
for effective energy density $\tilde\rho \tilde A$, in which also $\tilde A(G)$ depends only locally on the background. In a second step we will then be able to derive the equation of motion for the point particle fluid; this will turn out to be the equation for non-null Finsler geodesics.

We substitute the ansatz (\ref{sourceps}) for the particle string fluid source tensor into the source conservation equation. The result can be rewritten in the form
\begin{eqnarray}\label{cons2}
0 & = & |\textrm{Det }G|^{1/6}\textrm{det }\tilde g^{-1}  \tilde\rho \, \partial_p \tilde{\mathcal{G}}_{ijkl}u^iu^ju^ku^l - 4 \partial_l \Big( |\textrm{Det }G|^{1/6} \tilde\rho \frac{\mathcal{G}_{pijk}u^iu^ju^k}{\mathcal{G}(u,u,u,u)}u^l \Big)\nonumber\\
& & -|\textrm{Det }G|^{1/6} \tilde\rho \Sigma^{ijkl}\partial_p G_{ijkl}\\
& & +4\partial_l\left(|\textrm{Det }G|^{1/6} \tilde\rho \Sigma^{lijk} G_{pijk}\right) - 2\partial_p\left(|\textrm{Det }G|^{1/6} \tilde\rho\right)\,. \nonumber
\end{eqnarray}
Here $\tilde{\mathcal{G}}_{abcd}=|\textrm{Det }G^C|^{1/3}\mathcal{G}_{abcd}$. The derivation of this result requires the following identities, whose proof is rather technical, but can be performed with relative ease in the frame $\{e_{\hat 0},e_{\hat\alpha}\}$ with $e_{\hat 0}=u$:
\begin{eqnarray}
\tilde g^{-1\,aijk}G_{pijk} & = & -2\delta^a_p -4\frac{\mathcal{G}_{pijk}u^iu^ju^k}{\mathcal{G}(u,u,u,u)}u^a\,,\\
\frac{\delta\tilde{\mathcal{G}}(u,u,u,u)}{\delta G^C_{abcd}} & = & -\frac{1}{4}\textrm{det }\tilde g \tilde g^{-1\,abcd}\,.
\end{eqnarray}

A first restriction on the term $\Sigma$ can now be obtained by using the fact that it depends only locally on the background geometry: hence any condition on $\Sigma$ that is derived for constant area metric components $G_{abcd}$ must also hold for general backgrounds. We therefore set all partial derivatives of $G$ in (\ref{cons2}) to zero, and contract with $u^p$. This is the only possible scalar contraction, and so must imply the continuity equation~(\ref{continuity}). This requires
\begin{equation}\label{sigma1}
\Sigma^{lijk}G_{pijk}=\frac{1}{2}\delta^l_p\,.
\end{equation}
Substituting this into the full equation (\ref{cons2}), the last line is precisely cancelled. We again contract with $u^p$ to obtain a scalar equation, now for generic backgrounds $G$. To conveniently simplify the calculation we use the normalization $\mathcal{G}(u,u,u,u)=1=|\textrm{Det }G^C|^{-1/3}\textrm{det }\tilde g$ in terms of the totally symmetric tensor $\mathcal{G}$ defined in~(\ref{dualFresnel}). It is then straightforward to show that the continuity equation~(\ref{continuity}) can be obtained for interaction terms $\Sigma$ that also satisfy the condition
\begin{equation}\label{sigma2}
\Sigma^{ijkl}\partial_p G_{ijkl} = \partial_p\ln B\,,
\end{equation}
for some scalar density $B(G)$. The function $\tilde A$ is then determined by
$\tilde A=\tilde A_0 |\textrm{Det }G^C|^{-1/12}B$.

We now employ the two conditions (\ref{sigma1}) and (\ref{sigma2}) for $\Sigma$ in the source conservation equation~(\ref{cons2}) which yields the simplified equivalent expression
\begin{eqnarray}
0 & = & \partial_\tau (\mathcal{G}_{pijk}u^iu^ju^k) - \frac{1}{4}|\textrm{Det }G^C|^{-1/3}\partial_p\tilde{\mathcal{G}}_{ijkl}u^iu^ju^ku^l\nonumber\\
& & {}-\mathcal{G}_{pijk}u^iu^ju^k\partial_\tau\ln\big(\tilde A/\tilde A_0\big)+\partial_p\ln\big(|\textrm{Det }G^C|^{1/12}\tilde A/\tilde A_0\big)
\end{eqnarray}
where $\partial_\tau=u^p\partial_p$. Note that all dependence on the string fluid's energy density $\tilde\rho$ has cancelled. It is not hard to prove now that this equation can be equivalently derived as the stationarity equation from the point particle action
\begin{equation}\label{particleaction}
   \int d\tau\,\tilde A^{-4}\mathcal{G}(\dot x,\dot x,\dot x,\dot x)  
\end{equation}
together with the normalization constraint $\mathcal{G}(\dot x,\dot x,\dot x,\dot x)=1$.

Thus we have shown that all string fluids with the source tensor (\ref{sourceps}) and string interaction terms $\Sigma$ solving both conditions~(\ref{sigma1}) and~(\ref{sigma2}) behave as non-interacting particle fluids. The source conservation equation for these string fluids not only implies the particle fluid conservation equation (\ref{continuity}), but also an equation of motion for the fluid worldlines: the equation for non-null Finsler geodesics~\cite{Shenbook} with respect to the Finsler norm $\tilde A^{-1}(\mathcal{G}(\dot x,\dot x,\dot x,\dot x))^{1/4}$. This Finsler norm is fully determined by the area metric $G$ through its associated totally symmetric dual Fresnel tensor~$\mathcal{G}$, see definition (\ref{dualFresnel}). 

The continuity equation (\ref{continuity}) shows that the appearance of the function $\tilde A$ originates from a redefinition of the resulting particle fluid's energy density $\rho=\tilde\rho \tilde A/\tilde A_0$ in terms of the string fluid variable $\tilde\rho$. Mathematically, $\tilde A$ simply presents a conformal rescaling of the Finsler norm. This function can be fixed as $\tilde A/\tilde A_0=1$ by identifying the tension per area $\tilde\rho$ of the strings with the energy density $\rho$ of the effective point particles~\cite{Letelier:1984dm}.

%%%%%%%%%%%%%%%%%%%%%%%%%%%%%%%%%%%%%%%%%%%%%%%%%%%%%%%%%%%%%%%%%%%%%%
\section{Consistency}\label{sec_consistency}
Recall that a diffeomorphism invariant coupling of point particles to any gravity theory for a metric spacetime already determines the motion of these particles along Riemannian geodesics. This is because diffeomorphism invariance will result in some Bianchi identity for the gravitational curvature tensor and energy momentum conservation of the point particles. The latter requires the worldlines to follow geodesics. So given the same initial conditions, point particles will follow the same worldlines, independent of their masses. We will now explicitly demonstrate an analogous result for area metric spacetimes. Also here the motion of point particles along Finsler geodesics can be understood as a consequence of diffeomorphism invariance; thus these point particles can be consistently coupled to any area metric gravitational field equations. 

In the previous section we found that the point particle limit of a string fluid leads to the action~(\ref{particleaction}) with $\tilde A/\tilde A_0=1$. The corresponding source tensor is then
\begin{equation}\label{pointsource}
     T_{ijkl}(y) = -m \int d\tau \, \frac{\delta(y-x(\tau))}{|\textrm{Det G}|^{1/6}} \, \frac{\delta\mathcal{G}_{abcd}(y)}{\delta G^{ijkl}}\, \dot x^a\dot x^b\dot x^c\dot x^d  \,,   
\end{equation}
where $m$ denotes either the energy of a photon, or the mass of a massive particle. Since this source tensor is obtained from an action, it satisfies the conservation equation (\ref{conservation}). We will now show that this equation can be rewritten as
\begin{equation}\label{eq_finsler}
  - m \int d\tau \delta(y-x(\tau)) \left[ \partial_p \mathcal{G}_{abcd} \dot x^a\,\dot x^b\,\dot x^c\,\dot x^d\, + 4 \frac{d}{d\tau}(\mathcal{G}_{pabc}\, \dot x^a\,\dot x^b\,\dot x^c\, )  \right] = 0\,.
\end{equation}
Since this equation must hold for any point $y$, the expression in square brackets must vanish, which is precisely the equation of a geodesic in a Finsler geometry determined by the Fresnel tensor $\mathcal{G}$.

Indeed, the variation of the Fresnel tensor with respect to the inverse area metric, as it appears in the source tensor (\ref{pointsource}), can be expressed as a variation with respect to the cyclic part $G^C$ of the area metric~$G$ as
\begin{equation}\label{GCvariation}
  \frac{\delta \mathcal{G}_{abcd}}{\delta G^{ijkl}} = \frac{\delta \mathcal{G}_{abcd}}{\delta G^C_{\alpha\beta\gamma\delta}}\frac{\delta(G_{\alpha\beta\gamma\delta}-G_{[\alpha\beta\gamma\delta]})}{\delta G^{ijkl}}
 = -\frac{1}{4} \frac{\delta \mathcal{G}_{abcd}}{\delta G^C_{\alpha\beta\gamma\delta}} \left(G_{\alpha\beta ij}G_{\gamma\delta kl} - G_{ij[\alpha\beta}G_{\gamma\delta]kl}\right).
\end{equation}
The first term of the source conservation equation can now be evaluated as  
\begin{equation}\label{term1}
  \textrm{first term} \, = \, m \int d\tau\, \delta(y-x(\tau)) \frac{\delta \mathcal{G}_{abcd}}{\delta G^C_{\alpha\beta\gamma\delta}} \left(\delta^{rstu}_{\alpha\beta\gamma\delta} - \delta^{rstu}_{[\alpha\beta\gamma\delta]}\right) \partial_p G_{rstu} \dot x^a \dot x^b \dot x^c \dot x^d\,,
\end{equation}
which via the chain rule already yields the first term of the Finsler geodesic equation, in the square brackets of (\ref{eq_finsler}). The simplification of the second term of the source conservation equation is slightly more involved. First observe that
\begin{equation}
  \frac{\delta\mathcal{G}_{abcd}}{\delta G^C_{\alpha\beta\gamma\delta}} = -\frac{1}{12} (G^C)^{-1}{}^{\alpha\beta\gamma\delta} \mathcal{G}_{abcd} + |\textrm{Det G}|^{-1/3} \frac{\delta \tilde{\mathcal{G}}_{abcd}}{\delta G^C_{\alpha\beta\gamma\delta}} \,,
\end{equation}
where we defined the Fresnel tensor density $\tilde{\mathcal{G}}_{abcd} = |\textrm{Det} G^C|^{1/3} \mathcal{G}_{abcd}$. Using this the second term becomes
\begin{equation}\label{secondterm}
   -2m \int d\tau \frac{\partial}{\partial y^p} \delta(y-x(\tau)) \mathcal{G}_{abcd} \dot x^a \dot x^b \dot x^c \dot x^d
   + 4m \int d\tau \frac{\partial}{\partial y^\beta} \delta(y-x(\tau)) |\textrm{Det} G|^{-1/3} A^\beta{}_p\,,
\end{equation}
where we have introduced a shorthand for the quantity
\begin{equation}
  A^\beta{}_p = \frac{\delta \tilde{\mathcal{G}}_{abcd} \dot x^a \dot x^b \dot x^c \dot x^d}{\delta G^C_{\alpha\beta\gamma\delta}} G^C_{\gamma\delta\alpha p}\,.
\end{equation}
This quantity is most efficiently calculated using a non-holonomic frame $\{e_{\hat k}\}$ with $e_{\hat 0} = u$, and a dual frame $e^{\hat k}$. After considerable algebra one finally arrives at
\begin{equation}\label{Ainter}
  A^\beta{}_p = |\textrm{Det} G|^{1/3} \left(\mathcal{G}_{pabc} \dot x^a \, \dot x^b \,\dot x^c \,\dot x^\beta + \frac{1}{2} \delta^\beta_p \mathcal{G}_{abcd} \dot x^a \, \dot x^b \,\dot x^c \,\dot x^d\right)\,.
\end{equation}
Insertion of this result into (\ref{secondterm}) then provides the second term of equation~(\ref{eq_finsler}):
\begin{equation}\label{term2}
   \textrm{second term} \, = \, 4m \int d\tau \, \dot x^p\frac{\partial}{\partial y^p} \, \delta(y-x(\tau)) \, \mathcal{G}_{abcp} \dot x^a \dot x^b \dot x^c\,.
\end{equation}

Hence in any diffeomorphism invariant theory of area metric gravity, the field equations consistently couple to effective point particles propagating along Finsler geodesics. This coupling is universal for all point particles, irrespective of their mass (or energy for light). This in fact shows consistency with the experimentally supported weak equivalence principle.

%%%%%%%%%%%%%%%%%%%%%%%%%%%%%%%%%%%%%%%%%%%%%%%%%%%%%%%%%%%%%%%%%%%%%%%%%%%%
\section{Conclusions}\label{sec_discussion}
In this article, we have calculated the paths of massive point-like matter on general area metric manifolds. Even though point-like particles do not arise as fundamental mechanical objects on area metric backgrounds, their effective description is of phenomenological relevance.   

Intriguingly, we find that an area metric background impresses itself as a Finsler geometry on the motion of all point-like matter. Free motion is described by Finsler geodesics, and the relevant Finsler norm is determined by the area metric. To obtain this result, we have constructed the class of classical string fluids that admit a particle fluid limit through a geometrically well-defined averaging process. It turned out that the massive case considered here is governed by precisely the same Finsler geometry as the propagation of light~\cite{Punzi:2007di}. 
Taking into account recent studies of Finsler geometries in connection to the quantization of deformed general relativity~\cite{Girelli:2006fw,Vacaru:2007nh} and to quantum generalizations of the Poincar\'e algebra~\cite{Gibbons:2007iu}, it is interesting to note that the area metric structure of spacetime attaches a prominent role to a particular Finsler geometry when it comes to the description of the effective motion of light and matter. 

Of course, the Finsler geodesics found here simply reduce to the standard metric geodesics in case the area metric is induced by a metric. However, already $\omega=0$ Brans--Dicke theory determines a true area metric background with an additional degree of freedom, namely, the scalar field~\cite{Punzi:2008dv}. In this case the Finsler geodesics are the geodesics of a particular conformally rescaled metric, with the result that the theory is rendered consistent with solar system physics.

The general result that light and massive matter propagates along Finsler geodesics becomes inevitable if the area metric structure is taken seriously as the geometry of spacetime. Indeed, spacetime backgrounds described by a more general structure than metric geometry arise in various approaches to quantum gravity, and in string theory where additional massless background  fields appear. Especially important for the physical interpretation of generalized geometries are the classical tests of gravity in the solar system that require a model of planetary motion via distinguished curves. 
Similarly in cosmology, the trajectories of galaxies in an area metric spacetime must be understood as arising from a fluid model of the cosmological matter distribution. Our result identifies this mechanism.

Thus the findings of this article are essential in order to test the viability of the hypothesis of any generalized geometric backgrounds.

\acknowledgments
RP and MNRW acknowledge full financial support from the German Research Foundation DFG through the Emmy Noether Fellowship grant WO 1447/1-1.

%%%%%%%%%%%%%%%%%%%%%%%%%%%%%%%%%%%%%%%%%%%%%%%%%%%

\end{document}